\documentclass[twocolumn]{svjour3}
\usepackage{latexsym}
\usepackage{amsmath}
\usepackage[dvips]{graphicx}
\begin{document}
\title{Suppression of growth by multiplicative white noise in a parametric resonant system}
\author{Masamichi Ishihara} 
\institute{Koriyama Women's University, Department of Human Life Studies
  \email{m\_isihar@koriyama-kgc.ac.jp}
}
\date{}

\maketitle

\begin{abstract}
The author studied the growth of the amplitude 
in a Mathieu-like equation with multiplicative white noise.
The approximate value of the exponent at the extremum on parametric resonance regions 
was obtained theoretically by introducing the width of time interval, and 
the exponents were calculated numerically by solving the stochastic differential equations 
by a symplectic numerical method.
The Mathieu-like equation contains a parameter $\alpha$ that is determined by 
the intensity of noise and the strength of the coupling between the variable and the noise.
The value of $\alpha$ was restricted not to be negative without loss of generality.
It was shown that 
the exponent decreases with $\alpha$, reaches a minimum and 
increases after that. 
It was also found that 
the exponent as a function of $\alpha$ has only one minimum at $\alpha \neq 0$ 
on parametric resonance regions of $\alpha = 0$.
This minimum value is obtained theoretically and numerically.
The existence of the minimum at $\alpha \neq 0$ indicates 
the suppression of the growth by multiplicative white noise.
\end{abstract}

\section{Introduction}
In past few decades, many researchers have investigated the roles of noise, 
and then marked phenomena were found.
Such phenomena are 
stochastic resonance \cite{Gammaitoni,Collins,Yang,Tessone},
phase transition induced by multiplicative noise \cite{Broeck}, 
etc \cite{Chialvo,FUKUDA,Zaikin,Miyakawa,Pikovsky}.
A basic system in which multiplicative noise acts is an oscillator with varying mass. 
Oscillators in the presence of noise were investigated
\cite{Stratonovich,Landa,Mallick2002,Mallick2003,Mallick2005eprint}
and it was shown that the amplitude is amplified.

Another mechanism of growth is parametric resonance \cite{Landau}.
The effects of additive white noise 
acting on a harmonic oscillator with a periodic coefficient 
has been investigated \cite{Zerbe}.
Mean square displacement of an oscillator driven by a periodic coefficient 
was also studied in the presence of additive white noise \cite{Tashiro,Tashiro09}. 
The parametric resonance induced by multiplicative colored noise 
was investigated in ref.~\cite{Bobryk}. 
Experimentally, some physical systems which are described by the equations 
with a periodic coefficient and a multiplicative noise term were studied \cite{Berthet}.

A differential equation with a periodic coefficient and a multiplicative noise term 
appears in some systems. 
Multiplicative noise may amplify or suppress the amplitude, as additive noise does.
The magnitude of the amplitude is directly related to the stability of the system
and the physical quantities, such as energy and the number of particle. 
Thus the effects of multiplicative white noise should be investigated
in a parametric resonant system.

In this paper, 
a stochastic differential equation was analyzed by introducing the width of time interval
in a parametric resonant system.
The equation contains a parameter $\alpha$ that is determined by 
the intensity of noise and the strength of the coupling between the variable and the noise.
The value of $\alpha$ was restricted not to be negative in the equation, 
without loss of generality.
I estimated the exponent that indicates the growth of the amplitude.
I showed the existence of the minimum of the exponent and 
estimated the minimum value as a function of $\alpha$ 
by deriving an approximate expression of the exponent 
on parametric resonance regions of $\alpha=0$.
The stochastic differential equations were solved numerically by a symplectic method 
to avoid the growth by numerical error, 
and the exponent was extracted from the average of the trajectories.
The behavior of the exponent as a function of $\alpha$ was displayed numerically.

I found that 
the exponent 
has only one minimum at $\alpha \neq 0$ on parametric resonance regions of $\alpha = 0$ 
and that the relative variation is of the order of 90\%.
The existence of the minimum indicates the suppression of the growth 
by multiplicative white noise.
The results provide insight in the systems with periodically varying parameters and 
multiplicative noise. 
The multiplicative noise should suppress the growth in a parametric resonant system
when the intensity of noise and the coupling strength are appropriate.

\section{The exponent on parametric resonance regions}
\label{sec:exponent}
\subsection{An approximate equation of the exponent}
An equation with a periodic coefficient and a multiplicative white noise term 
is interested in some branches of physics
\cite{Berthet,Zanchin,Ishihara7}.
A typical equation is
\begin{equation}
\ddot{\phi} + \left[ 1 + \beta \cos \left( \gamma z \right) + \alpha r(z) \right] \phi = 0,
\label{eqn:start_equation}
\end{equation}
where the dot represents the derivative with respect to $z$.
The quantity $r(z)$ has the following properties:
\begin{equation}
\langle r(z) \rangle = 0,  \qquad \langle r(z) r(z') \rangle = \delta(z-z') ,
\label{eqn:properties_of_r}
\end{equation}
where the notation $\langle \cdots \rangle$ represents statistical average.
The value of $\alpha$ is restricted not to be negative in Eq.~\eqref{eqn:start_equation}
without loss of generality.
The starting point in this study is 
Eq.~\eqref{eqn:start_equation} with Eq.~\eqref{eqn:properties_of_r}.

Equation \eqref{eqn:start_equation} is rewritten with the variable $p_{\phi}$ 
which is defined by $p_{\phi} = d\phi/dz$:
\begin{subequations}
\begin{align}
d\phi     &= p_{\phi} dz, 
\label{eqn:numerical_phi} \\
dp_{\phi} &= - \left[ 1 + \beta \cos \left(\gamma z \right) \right] \phi dz 
             - \alpha \phi \circ dW.
\label{eqn:numerical_p_phi} 
\end{align}
\end{subequations}
The quantity $W(z)$ is defined by $W(z) = \int_{z_0}^{z} ds \ r(s) $ 
and this is a wiener process,
where the quantity $z_{0}$ is an initial time. 
(Here, the symbol $\circ$ represents Stratonovich product.)
I attempt to solve Eqs.~\eqref{eqn:numerical_phi} and \eqref{eqn:numerical_p_phi} numerically 
in \S~\ref{sec:numerical_calculation}. 

Equation~\eqref{eqn:start_equation} is just a Mathieu equation when $\alpha$ is zero,
and this equation has resonance bands. 
With the relation $2u=\gamma z$, 
the Mathieu equation corresponding to Eq.~\eqref{eqn:start_equation} is given by
\begin{equation}
\frac{d^{2}\phi}{du^{2}} + 
\left( a  - 2q  \cos(2u) \right) \phi =0,
\label{eqn:mathieu_equation}
\end{equation}
where $a = {4}/{\gamma^{2}}$ and $-2q ={4 \beta}/{\gamma^{2}}$.
Then the bands are distinguished by positive integer $n$ 
with the relation $n^{2} = {4}/{\gamma^2}$ .
Therefore the values of $\gamma$ in resonance bands at $\alpha = 0$ are close to $2/n$.

In this paper, I attempt to estimate the growth rate of the amplitude in time.
This rate is obtained from the exponent which is given by 
${\displaystyle \mathop{\lim\sup}
\ z^{-1} \ln \left[\left|\langle \phi(z) \rangle \right| / \left|\phi_0 \right|\right]}$
, where $\phi_0$ is the initial value.
I use the solution of the Mathieu equation 
to solve Eq.~\eqref{eqn:start_equation} approximately 
in the resonance regions of $\alpha = 0$.
The equation at $\alpha=0$ is 
\begin{equation}
\ddot{\Phi} + \left[ 1 + \beta \cos \left(\gamma z \right) \right] \Phi = 0,
\label{eqn:mathieu:Phi}
\end{equation}
The quantity $\phi$ is represented as a product of $\Phi$ multiplied by a new variable $\psi$:
$\phi = \Phi \psi$.
The quantity $\psi$ satisfies the subsequent equation:
\begin{equation}
\ddot\psi + 2 \left( {\dot{\Phi}}/{\Phi} \right) \dot{\psi} +  \alpha r(z) \psi  = 0. 
\label{eqn:psi}
\end{equation}
The exponent of $\Phi$ was investigated by many researchers in detail. 
Thus, the exponent of $\phi$ is estimated by obtaining the exponent of $\psi$ approximately.

Here I denote the exponent of $\phi$ at $\alpha=0$ as $s \equiv s(\beta,\gamma)$ 
which is just the exponent of $\Phi$.
The time dependence of $\Phi$ is obtained by solving Eq.~\eqref{eqn:mathieu:Phi}.
One method to solve approximately in the 
resonance band 
is performed by putting the form of $\Phi$ with the assumption $\ddot{P}_n \sim 0$
as follows:\cite{Landau,Ishihara_Nonlinear,Son,Takimoto}
\begin{equation}
\Phi = 
\sum_{n=1} \left[ P_{n}(z) e^{i n \gamma z /2} + P_{n}^{*}(z) e^{-i n \gamma z /2} \right]
 + R(z).
\label{eqn:mode_decomposition}
\end{equation}
The growth of the function $P_{m}(z)$ is largest in the $m$th resonance band. 
Therefore, $\Phi$ in the $m$th band is approximately given by
\begin{subequations}
\begin{align}
& \Phi \sim e^{s_{m} z} F_{m}(z) , 
\label{eqn:s:m-th_band}\\
& F_{m}(z) := C e^{i m \gamma z /2} + C^{*} e^{-i m \gamma z /2} ,
\label{eqn:m-th_band}
\end{align}
\label{eqn:all:m-th_band}
\end{subequations}
where $C$ is a complex constant and $s_{m}$ is the exponent. 
It is conjectured that the exponent $s_{m}$ is close to the exponent $s$ 
in the $m$th resonance band. 
With Eqs.~\eqref{eqn:s:m-th_band} and \eqref{eqn:m-th_band}, I obtain
\begin{equation}
{\dot{\Phi}}/{\Phi} \sim s_{m} + {\dot{F}_{m}}/{F_{m}}.
\label{eqn:dotPhi_Phi}
\end{equation}
The exponent is estimated by solving Eq.~\eqref{eqn:psi} with Eq.~\eqref{eqn:dotPhi_Phi}. 
However, it is not easy to handle Eq.~\eqref{eqn:psi}. 
Instead, in Eq.~\eqref{eqn:psi}, 
I replace $({\dot{\Phi}}/{\Phi})$ by the average of $(\dot{\Phi}/\Phi)$ in time.
The average of $\dot{\Phi}/\Phi$ in one period of $F_{m}(z)$ is equal to $s_{m}$. 
Therefore, the approximate equation for $\psi$ under this approximation 
in the $m$th resonance band is 
\begin{equation}
\ddot{\psi} + 2 s_{m} \dot{\psi} + \alpha r(z) \psi = 0.
\label{eqn:approximated_equation_for_psi}
\end{equation}

To estimate the exponent, I put the form of $\psi$ as follows:
\begin{equation}
\psi = \psi_{0} \exp \left( \int_{z_0}^{z} dz' \sigma(z')  \right).
\label{eqn:psi_exp_form}
\end{equation}
Substituting Eq.~\eqref{eqn:psi_exp_form} into Eq.~\eqref{eqn:approximated_equation_for_psi},
I obtain the equation for $\sigma$:
\begin{equation}
\dot{\sigma} + \sigma^{2} + 2s \sigma + \alpha r(z) = 0,
\label{eqn:sigma}
\end{equation}
where the subscript $m$ of $s_m$ is omitted. 
In the next subsection, 
the exponent is obtained 
by estimating the statistical average $\langle \sigma \rangle$ approximately.

\subsection{The value of the exponent at the extremum on parametric resonance regions}
\label{subsec:Exponents_in_resonance_bands}
In this subsection, I estimate the minimum value of the exponent of $\phi$.
It is assumed that $r(z)$ is constant in the quite small time interval 
to estimate $\sigma$ given by Eq.~\eqref{eqn:sigma}.
The statistical average with respect to $r(z)$ is taken, because $r(z)$ varies randomly. 
The exponent of $\phi$ is estimated with the exponent $s$ and $\langle \sigma \rangle$.
The existence of the extremum of the exponent is obtained by differentiating 
the exponent with respect to $\alpha$.

At first, I find the solution when $r(z)$ is constant.
The solution of Eq.~\eqref{eqn:sigma} is categorized by the quantity $\cal{D}$
which is defined as $4s^{2}-4\alpha r$.
I have
\begin{align}
&
\int_{z_{0}}^{z} dz' \sigma(z') 
= 
\nonumber \\ & 
\left\{
\begin{array}{ll}
\left( -s+ \frac{\sqrt{\cal{D}}}{2} \right)  \left(z-z_{0}\right)
+ 
\ln \left| \frac{1-Ce^{-\sqrt{\cal{D}} z }}{1-Ce^{-\sqrt{\cal{D}} z_{0} }} \right|
& \quad {\cal D} > 0 \\
- s \left(z-z_{0}\right) + \ln \left| \frac{z+C'}{z_{0}+C'} \right|
& \quad {\cal D} = 0 \\
 - s \left(z-z_{0}\right)
  + \ln \left| 
\frac{\cos\left(\frac{\sqrt{-{\cal D}}}{2} z_{0} + C''\right)}
{\cos\left(\frac{\sqrt{-{\cal D}}}{2} z + C''\right)}\right|
& \quad {\cal D} < 0 
\end{array}
\right.
,
\label{eqn:sigma_integration}
\end{align}
where $C$, $C'$ and $C''$ are constants which are related to $\sigma(z_0)$.
The logarithm terms of the right-hand side in Eq.~\eqref{eqn:sigma_integration}
do not contribute to the growth substantially. 

Next, I treat the case that the quantity $r(z)$ is time dependent.
For such the case, the region $[z_{0},z]$ is divided 
into small regions of time interval $\Delta z$.   
Moreover, the region of the width $\Delta z$ is divided into quite small $N$ regions
numbered '$j$' in which the quantity $r$ is constant. 
I define the quantity $\Delta W_{j}$ by $r_{j} \Delta z / N$, 
where $r_{j}$ is the value of $r$ in the region '$j$'.
This quantity $\Delta W_{j}$ is a wiener process and 
the distribution function of $\Delta W_{j}$ is given by 
\begin{equation}
P(\Delta W_{j}) = 
\frac{1}{\sqrt{2\pi (\Delta z)/N}} 
\exp \left(- \frac{(\Delta W_{j})^{2}}{2 (\Delta z)/N} \right).
\end{equation}
Then the quantity $\Delta W \equiv \displaystyle\sum_{j=1}^{N} \Delta W_{j}$ 
obeys the distribution function $P(\Delta W)$ which is given by 
\begin{equation}
P(\Delta W) = 
\frac{1}{\sqrt{2\pi (\Delta z)}} 
\exp \left(- \frac{(\Delta W)^{2}}{2 (\Delta z)} \right).
\end{equation}
Therefore, 
the values of $\Delta W$ in the regions of time interval $\Delta z$ are distributed
with the probability $P(\Delta W)$. 
The statistical average of a variable ${\cal O}$ is given by
$\langle {\cal O} \rangle = \int_{-\infty}^{\infty} d(\Delta W) P(\Delta W) \ {\cal O}$.
From Eqs.~\eqref{eqn:s:m-th_band} and  \eqref{eqn:sigma_integration},
the exponent of $\phi$ in unit time of $z$ (I denote ${\cal G}$) should be estimated by
\begin{equation}
{\cal G} = \int_{-\infty}^{\infty} d(\Delta W) \ P(\Delta W) \Theta({\cal D}) 
\ \frac{\sqrt{{\cal D}}}{2},
\end{equation}
where 
$\Theta(x)$ is the step function which is 1 for $x > 0$ and 0 for $x<0$
, and ${\cal D}= 4s^{2} - 4 \alpha (\Delta W)/\Delta z$.
This integration can be performed and I obtain the following expression of ${\cal G}$:
\begin{equation}
{\cal G} = \frac{s}{2^{3/2} \kappa}
\exp \left( - \frac{\kappa^4}{4} \right) 
D_{-3/2} \left( - \kappa^2 \right),
\label{eqn:res:exponent}
\end{equation}
where the variable $\kappa$ is defined as $(\Delta z)^{1/4} s / \alpha^{1/2}$ 
and $D_{\nu}$ is the parabolic cylinder function \cite{Abramowitz,Gradshteyn}.
The value ${\cal  G}/s$ depends only on $\kappa$,
and then the parameter $\Delta z$ affects ${\cal  G}/s$ through $\kappa$.
A certain value $\kappa$ is realized by adjusting $\alpha$ 
when $\Delta z$ is given. 
I can read the global behavior of the exponent as a function of $\alpha$ 
from Eq.~\eqref{eqn:res:exponent}.
The parameter $\alpha$ affects ${\cal G}/s$ through $\kappa$.

The quantity ${\cal G}$ as a function of $\alpha$ has an extremum
which is determined by $d{\cal G}/d\alpha$. 
I obtain the subsequent condition that ${\cal G}$ is extremum: 
\begin{equation}
D_{1/2} \left( - \kappa^2 \right) = 0.
\label{eqn:condition_of_minimum}
\end{equation}
It is known that $D_{\nu}(x)$ for positive $\nu$ has $[\nu+1]$ zeros \cite{Bateman},
where $[\nu+1]$ is the maximum integer which is not greater than $(\nu+1)$. 
Then the equation, $D_{1/2}(x)=0$, has one solution,
and I write the solution as $x_{\mathrm{sol}}$.
The value $x_{\mathrm{sol}}$ is negative, 
and then $\alpha$ is positive at the extremum of ${\cal G}$.
Therefore ${\cal G}$ has one extremum surely at a positive $\alpha$.
The value at the extremum of the exponent ${\cal G}$ is given by 
\begin{equation}
{\cal G}_{\mathrm{min}} = 
\frac{s}{2^{3/2}} \frac{1}{\left[-x_{\mathrm{sol}}\right]^{1/2}}
\exp \left( -\frac{1}{4} \left( x_{\mathrm{sol}} \right)^{2} \right)
D_{-3/2} \left( x_{\mathrm{sol}} \right) .
\label{eqn:min_val_of_G}
\end{equation}
The value ${\cal G}_{\mathrm{min}}$ is smaller than $s$.
That is,  ${\cal G}$ has one minimum at a positive $\kappa^2$.
This indicates that the exponent ${\cal G}$ is suppressed by multiplicative white noise 
when the value of $\alpha$ is appropriate.
I must note that the expression ${\cal G}_{\mathrm{min}}$ is independent of $\Delta z$.
Equation~\eqref{eqn:res:exponent} for quite small $s$ should be invalid, 
because the approximation of $\Phi$ given in Eq.~\eqref{eqn:all:m-th_band} does not work well.

\section{Numerical calculation of the exponents by a symplectic method}
\label{sec:numerical_calculation}
In this section, 
I attempt to solve Eqs.~\eqref{eqn:numerical_phi} and \eqref{eqn:numerical_p_phi} numerically. 
Our purpose is to obtain the amplitude of $\phi$ when white noise acts multiplicatively. 
Therefore, the amplitude must be calculated precisely,
at least, when a periodic coefficient and a white noise term are absent.
The system has the symplectic structure 
even when noise exists if some conditions are satisfied \cite{Milstein_additive}.
Taking this property into account, I use the symplectic method 
developed in ref.~\cite{Milstein_multiplicative} 
to solve the stochastic differential equations with multiplicative white noise.  
The first-order method given in ref.~\cite{Milstein_multiplicative} 
is applied to the equations in this study.
The equations are solved numerically from $z=0$ to $z=500$.
The time step in $z$ is set to 0.05.
The initial conditions are $\phi(0) = 1$ and $\dot{\phi}(0) = 0$ in these calculations. 

One trajectory of $\phi(z)$ can be calculated when one sequence of noise is given.
I calculate many trajectories and take their average 
to obtain the mean value of the trajectories of the variable $\phi_{i}^{(j)}(z)$,
where the subscript $i$ indicates the batch and the superscript $(j)$ indicates 
the trajectory  in a certain batch $i$.
In the present calculation, 
one batch contains 500 trajectories and 20 batches are used.
I calculate the mean value ${\cal M}_{i}(z)$ of the trajectories in the batch $i$. 
The mean value over 20 batches, $\bar{\phi}(z)$, is given by 
\begin{equation}
\bar{\phi}(z) = \frac{1}{20} \sum_{i=1}^{20} {\cal M}_{i}(z), 
\quad
{\cal M}_{i}(z) = \frac{1}{500} \sum_{j=1}^{500} \phi_{i}^{(j)}(z). 
\end{equation}
It is possible to perform interval estimation by using $\bar{\phi}$ and ${\cal M}_{i}$.
In the case of $\alpha = 0$, there is no need to calculate many trajectories. 
Thus only one trajectory is calculated numerically for $\alpha=0$.

The exponent is estimated from the average $\bar{\phi}(z)$ in the range of  $200<z<500$ 
to decrease the effects of the initial conditions. 
This estimation is performed as follows:
1) the sets $(z_{k},\ln\bar{\phi}(z_{k}))$ are determined, where $z_k$ is the time at which
$\bar{\phi}(z_{k})$ is a local maximum and positive.
2) the sets are fit with a linear function. 
The coefficient of the time $z$ is adopted as the exponent. 

Here, I note the reason why the values, $\ln\bar{\phi}(z_{k})$, are fit.
One way to estimate the parameters is to fit the average $\bar{\phi}(z_{k})$ directly. 
In such the method, it is implicitly assumed that 
the dispersion of the distribution of the data at time $z$ 
and that at time $z'$ ($\neq z$) are the same (approximately). 
However, the dispersion is wider with time $z$ in the present case, 
because I treat a wiener process.
The effects of non-equivalent dispersions are decreased by taking the logarithm of the data.
Therefore the transformed data, $\ln\bar{\phi}(z_{k})$, 
are fit with the linear function.
I notice that the exponents extracted by the above procedure are different generally 
from the Lyapunov exponents which 
are estimated by the mean value of the logarithm of $\phi_i^{(j)}$.
The quantity, $\ln\bar{\phi}(z_{k})$, is calculated,
because I focus on the enhancement of the variable $\phi$ in this study.

\begin{figure*}
\includegraphics[width=\textwidth]{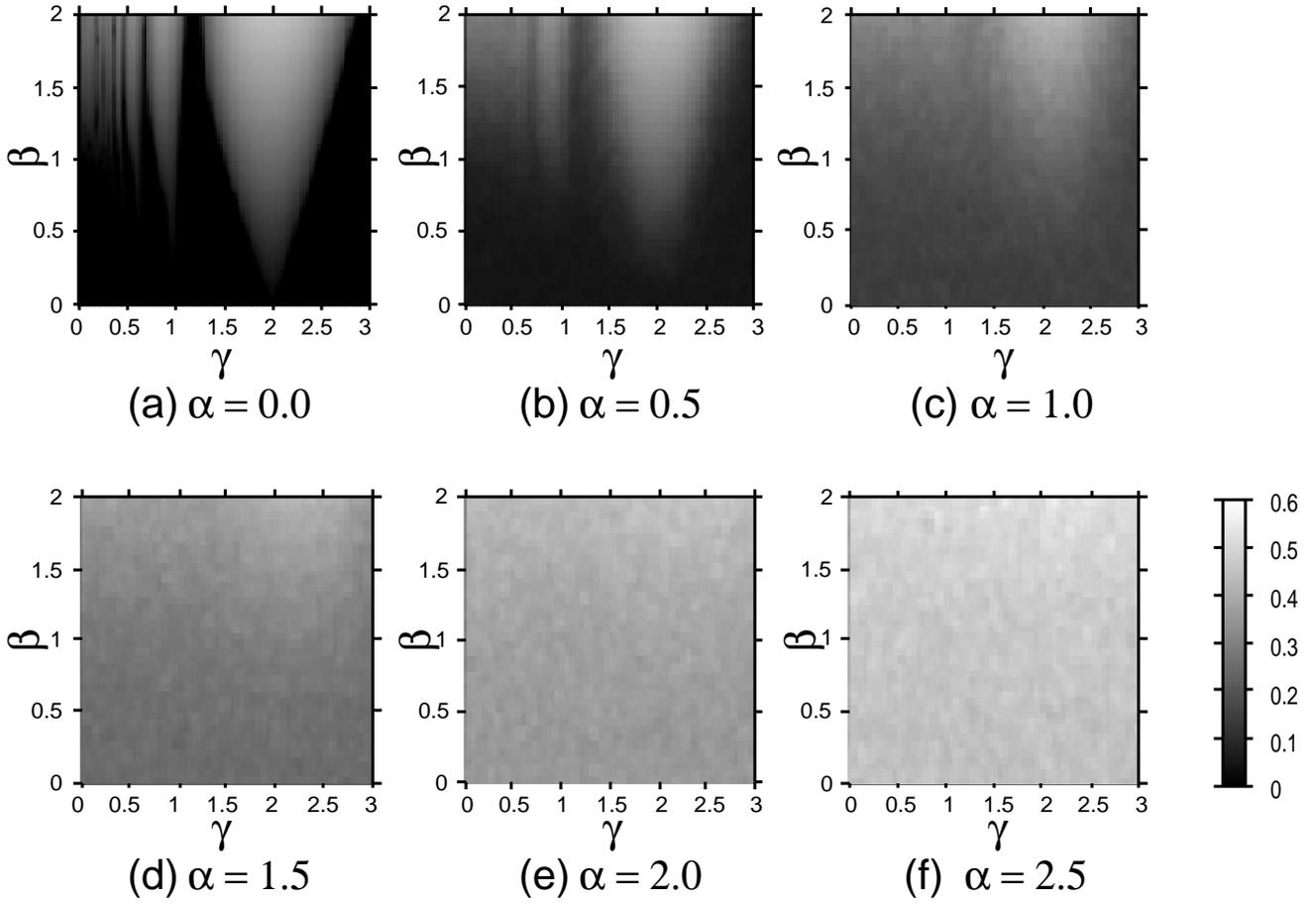}
\caption{
Exponents on the $\gamma$--$\beta$ plane for various values of $\alpha$.
The exponents are calculated  by solving the stochastic differential equations numerically
by the symplectic method.
The parameters are
(a)$\alpha=0.0$, (b)$\alpha=0.5$, (c)$\alpha=1.0$, 
(d)$\alpha=1.5$, (e)$\alpha=2.0$, (f)$\alpha=2.5$ respectively.
}
\label{figs:exponent_for_various_alpha}
\end{figure*}

\begin{figure*}
\begin{tabular}{cc}
\includegraphics[width=0.48\textwidth]{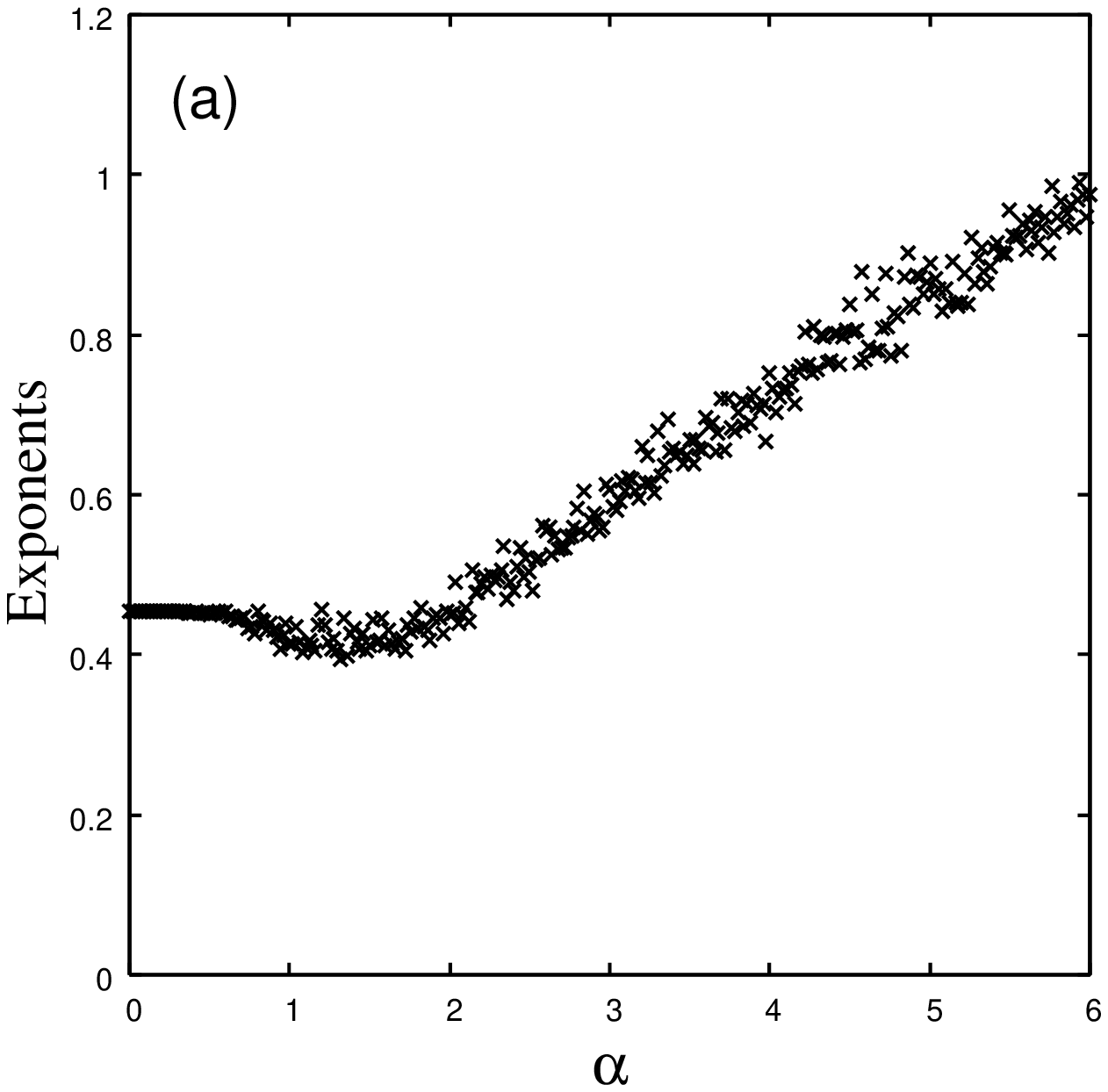}
& 
\includegraphics[width=0.48\textwidth]{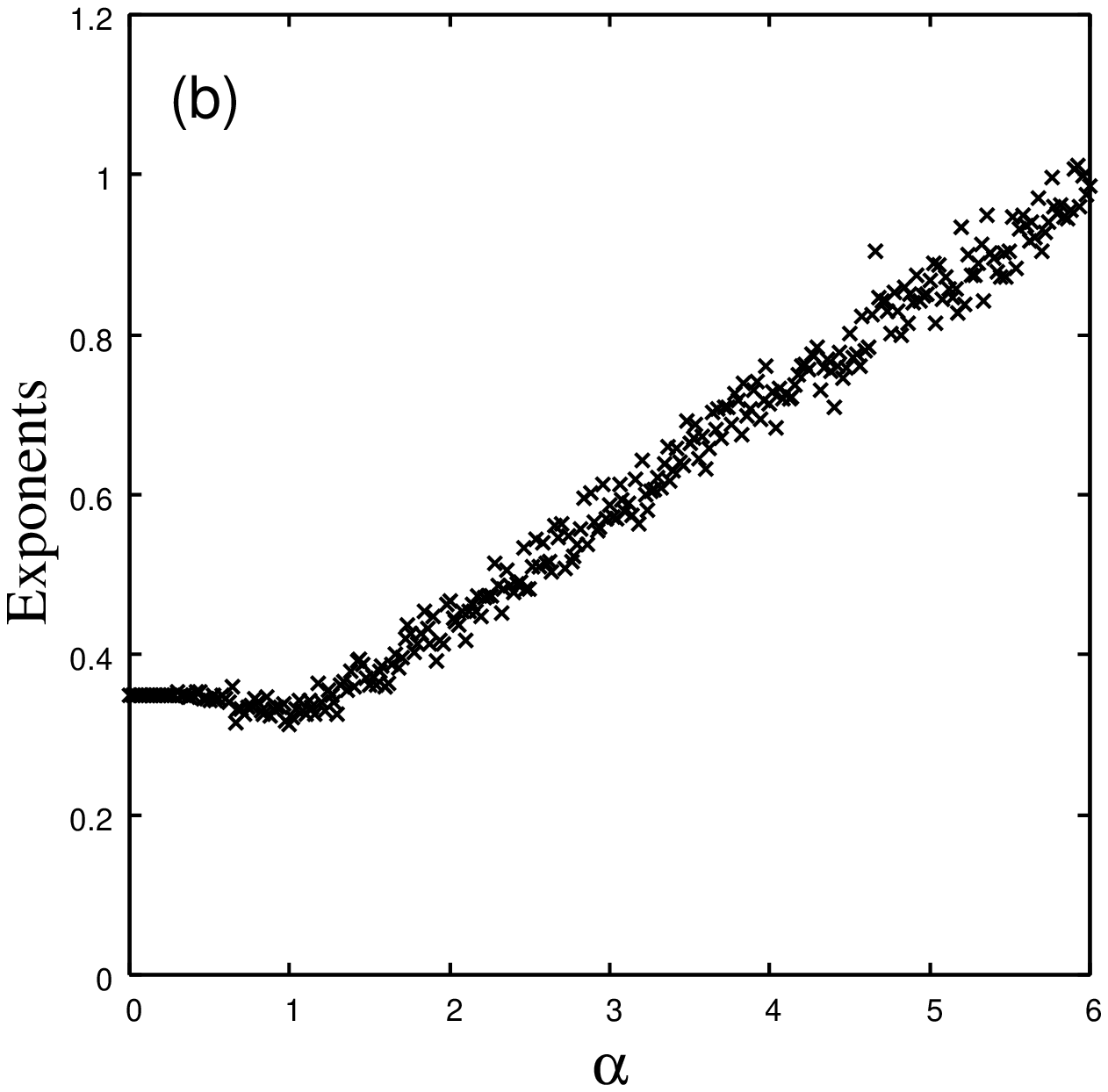}
\end{tabular}
\caption{
Exponents on the resonance bands.
The cross represents the data estimated numerically by solving 
the Eqs.~\eqref{eqn:numerical_phi} and \eqref{eqn:numerical_p_phi}.
(a) The values of the parameters, $\beta$ and $\gamma$, are both 2. 
(b) The values of the parameters, $\beta$ and $\gamma$, are 2 and 0.9 respectively. 
}
\label{figs::resonance:exponents:}
\end{figure*}

Figure~\ref{figs:exponent_for_various_alpha}(a) is the map 
of the exponents of the Mathieu equation, Eq.~\eqref{eqn:mathieu:Phi}, 
on the $\gamma$--$\beta$ plane.
The step sizes in $\gamma$ and $\beta$ in the numerical calculations are taken to be 0.02 
to draw this figure.
I denote these step sizes as  $\Delta \gamma$ and $\Delta \beta$ respectively.
The color of a square is determined from the arithmetic mean of the exponents 
at four corners which are located at
($\gamma$,$\beta$), ($\gamma$ + $\Delta \gamma$,$\beta$),
($\gamma$,$\beta$ + $\Delta \beta$) and  
($\gamma$ + $\Delta \gamma$,$\beta$ + $\Delta \beta$).
The resonance band around $\gamma=2$ corresponds to the first resonance band of 
Eq.~\eqref{eqn:mathieu_equation}.
The $n$th resonance band of Eq.~\eqref{eqn:mathieu_equation} corresponds to 
the band around $\gamma = 2/n$, where $n$ is positive integer.

Next, I show the map of the exponents for various values of $\alpha$ on the $\gamma$--$\beta$ plane.
Figure~\ref{figs:exponent_for_various_alpha}(b) is the map at $\alpha=0.5$, 
\ref{figs:exponent_for_various_alpha}(c) is at $\alpha=1.0$, 
\ref{figs:exponent_for_various_alpha}(d) is at $\alpha=1.5$, 
\ref{figs:exponent_for_various_alpha}(e) is at $\alpha=2.0$, 
and \ref{figs:exponent_for_various_alpha}(f) is at $\alpha=2.5$.
The step sizes in $\beta$ and $\gamma$ are 0.05  in the numerical calculations 
for Figs.~\ref{figs:exponent_for_various_alpha}(b), \ref{figs:exponent_for_various_alpha}(c),
 \ref{figs:exponent_for_various_alpha}(d), \ref{figs:exponent_for_various_alpha}(e) and 
 \ref{figs:exponent_for_various_alpha}(f).  
The color of a square is determined 
in the same manner as in Fig.~\ref{figs:exponent_for_various_alpha}(a). 
As shown in Figs.~\ref{figs:exponent_for_various_alpha}(b), 
\ref{figs:exponent_for_various_alpha}(c), \ref{figs:exponent_for_various_alpha}(d), 
\ref{figs:exponent_for_various_alpha}(e) and \ref{figs:exponent_for_various_alpha}(f), 
the band structure is destroyed by noise, 
and the values of the exponents become large with $\alpha$ for many sets of $(\gamma,\beta)$. 
However it seems from these figures that 
the exponent on the resonance band is not a monotonically increasing function of $\alpha$.
Moreover, 
the $\beta$ dependence of the exponent in Fig.~\ref{figs:exponent_for_various_alpha}(f) is weak 
as compared with those in other figures:
Figs.~\ref{figs:exponent_for_various_alpha}(a), 
\ref{figs:exponent_for_various_alpha}(b) and \ref{figs:exponent_for_various_alpha}(c).
This fact in Fig.~\ref{figs:exponent_for_various_alpha}(f) implies that 
the values of the exponents of the equation with the periodic coefficient are close to 
those without the periodic coefficient. 
(The values of the exponents at $\beta=0$ correspond to 
the values in the case of no periodic coefficient.)
It is evident that the effects of the periodic coefficient become weak relatively.

Furthermore,
I investigate the $\alpha$ dependence of the exponent 
on the first and the second resonance bands.
I draw the $\alpha$ dependence of the exponent with the fixed parameters, $\gamma$ and $\beta$.
I show the exponents for the set $(\gamma=2,\beta=2)$ on the first resonance band, 
and the set $(\gamma=0.9,\beta=2)$ on the second resonance band.
Figure~\ref{figs::resonance:exponents:} shows 
the $\alpha$ dependences of the exponents. 
The cross represents the data obtained by solving 
Eqs.~\eqref{eqn:numerical_phi} and \eqref{eqn:numerical_p_phi} numerically.
The suppression by noise is clearly seen and 
there is only one local minimum in each figure. 
The exponent decreases with $\alpha$ and reaches the minimum. 
It continues to increase with $\alpha$ after that.
This behavior is interpreted as follows. 
The growth of the amplitude depends on the mechanism of 
parametric resonance for small $\alpha$. 
This mechanism is destroyed by noise with the increase of $\alpha$. 
Then the exponent decreases with $\alpha$.
Contrarily, the amplitude is amplified by noise for large $\alpha$,
as shown in many researches. 
In summary,
the exponent decreases with $\alpha$, reaches the minimum, and increases after that. 
The exponents for other parameter sets, $(\gamma, \beta)$, on the resonance bands 
behave similarly.

Finally, 
the minimum value of the exponent as a function of $\alpha$ is estimated
for various values of $\beta$ and $\gamma$.
I denote the minimum value of the exponent estimated numerically as $s_{\mathrm{min}}$. 
Clearly $s_{\mathrm{min}}$ is a function of $\beta$ and $\gamma$.
In these calculations, the range of $\alpha$ is set to $[0,2]$ and 
the step size in $\alpha$ is set to 0.01.
The range of $\gamma$ is set to $[0.7,2.7]$ and the step size in $\gamma$ is set to 0.5.
The exponents for various values of $\alpha$ with the fixed $\beta$ and $\gamma$ are estimated 
and $s_{\mathrm{min}}$ is set to the minimum value of these exponents.
I calculate the quantity $s_{\mathrm{min}}/s$, 
because the exponent at $\alpha=0$, $s$, is also a function of $\beta$ and $\gamma$.
I show the values $s_{\mathrm{min}}/s$ for $s \ge 0.3$ 
to compare them with the value ${\cal G}_{\mathrm{min}}/s$.
The value ${\cal G}_{\mathrm{min}}/s$ is approximately 0.893 from Eq.~\eqref{eqn:min_val_of_G}.

\begin{figure*}
\begin{tabular}{cc}

\includegraphics[height=0.43\textwidth, width=0.48\textwidth]{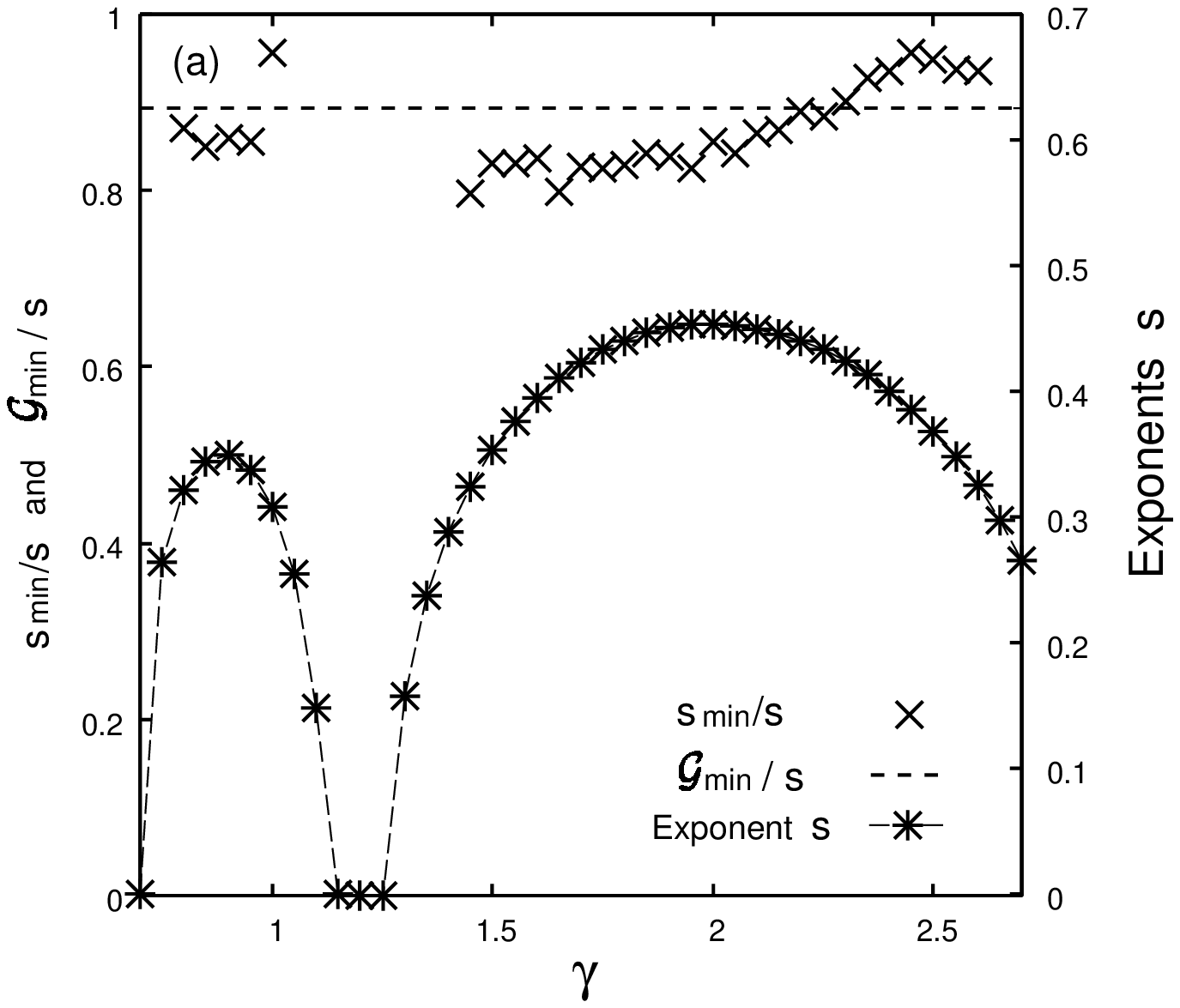}
& 
\includegraphics[height=0.43\textwidth,width=0.48\textwidth]{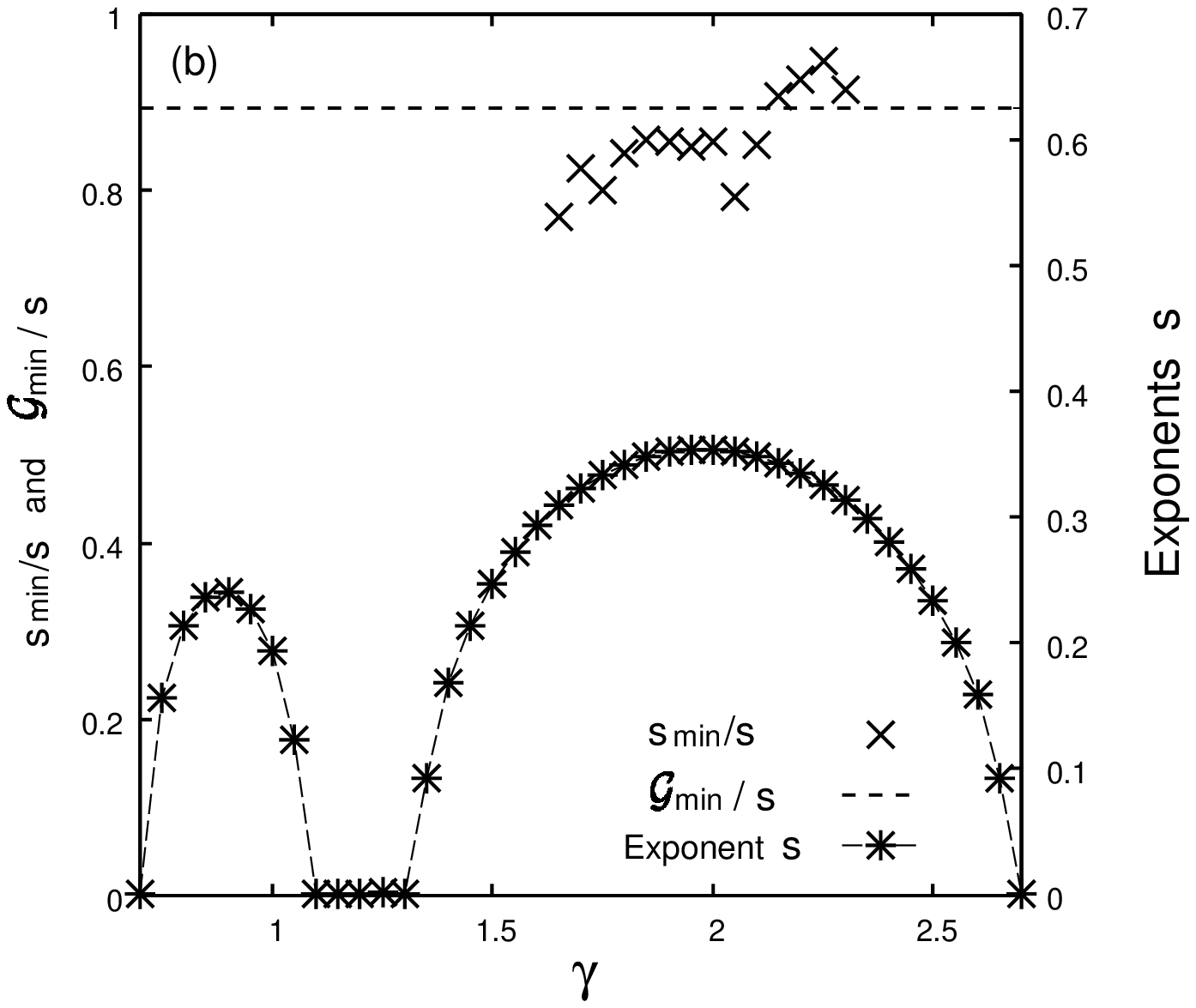}
\end{tabular}
\caption{
The values $s_{\mathrm{min}}/s$ for various values of $\gamma$. 
The parameter $\beta$ is set to (a) $\beta=2.0$ and (b) $\beta=1.5$.
Cross represents the data points of $s_{\mathrm{min}}/s$ for $s \ge 0.3$ and 
asterisk represents the data points of $s$. 
Broken line is the value ${\cal G}_{\mathrm{min}}/s$ which is approximately 0.893.
}
\label{figs:ratio:SMA1}
\end{figure*}

\begin{figure*}
\begin{tabular}{cc}

\includegraphics[height=0.43\textwidth, width=0.48\textwidth]{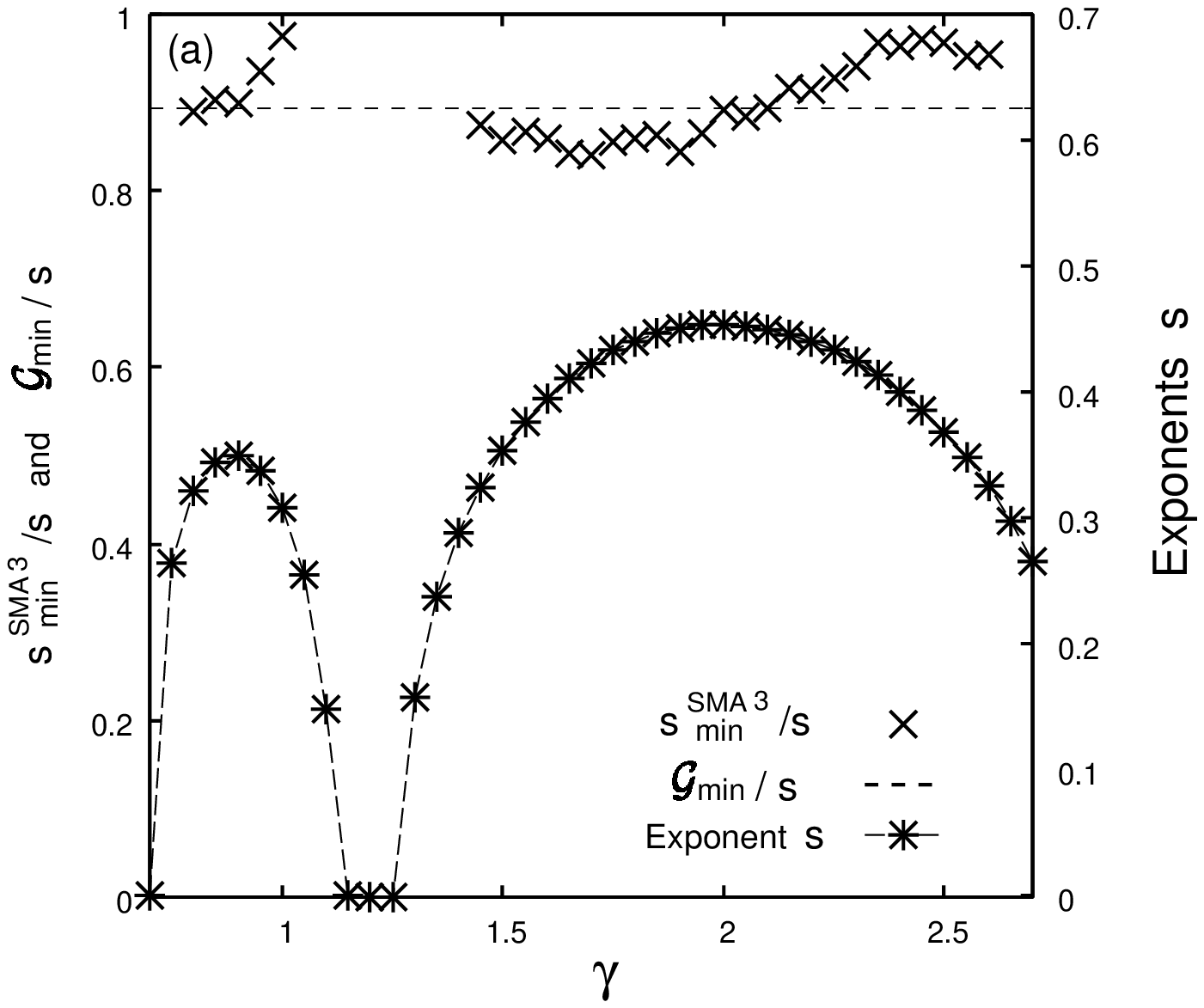}
& 
\includegraphics[height=0.43\textwidth,width=0.48\textwidth]{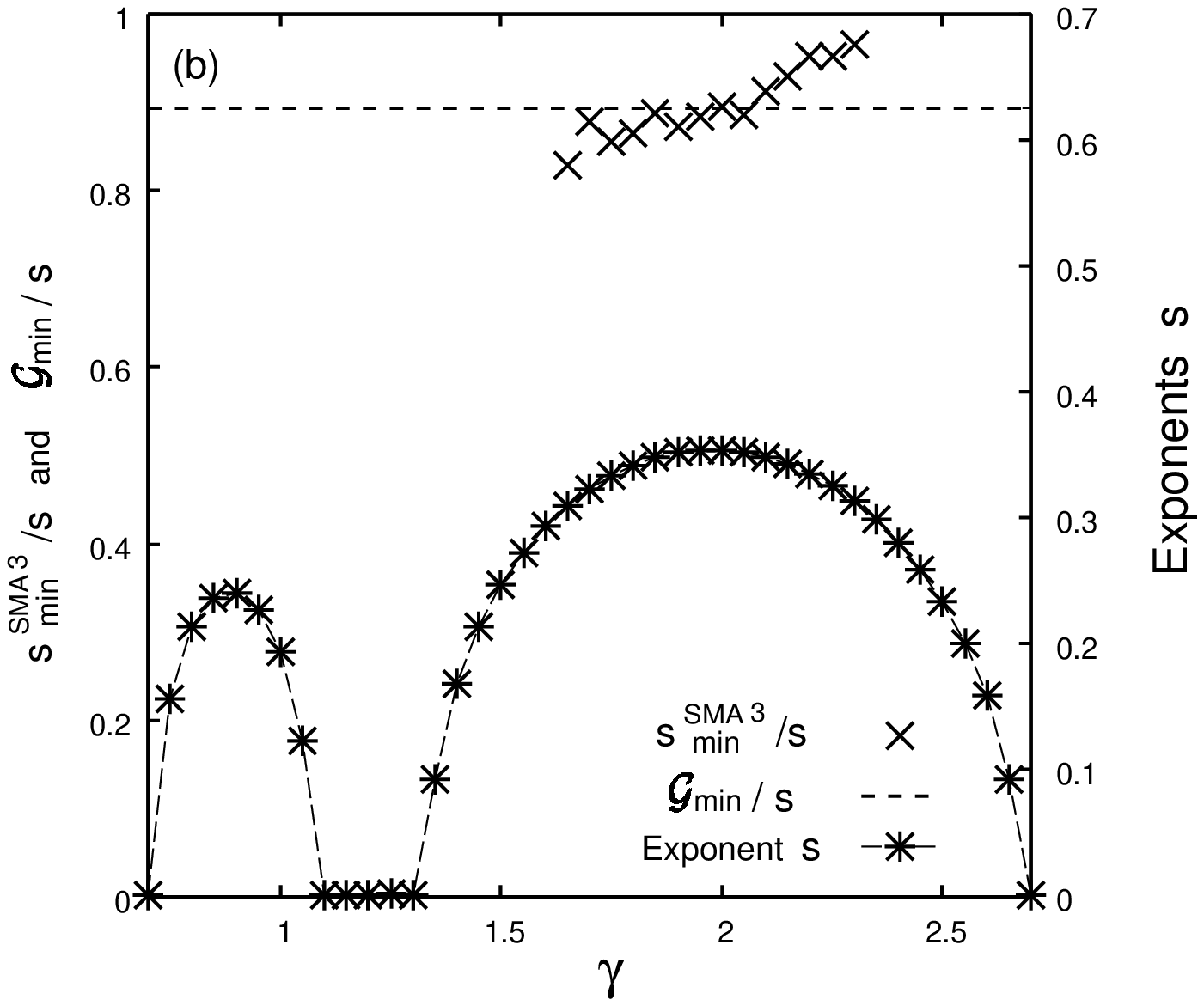}
\end{tabular}
\caption{
The values $s_{\mathrm{min}}^{\mathrm{SMA3}}/s$ for various values of $\gamma$. 
The parameter $\beta$ is set to (a) $\beta=2.0$ and (b) $\beta=1.5$.
Cross represents the data points of $s_{\mathrm{min}}/s$ for $s \ge 0.3$ and 
asterisk represents the data points of $s$. 
Broken line is the value ${\cal G}_{\mathrm{min}}/s$ which is approximately 0.893.
}
\label{figs:ratio:SMA3}
\end{figure*}

Figure~\ref{figs:ratio:SMA1} shows the values $s_{\mathrm{min}}/s$ for $s \ge 0.3$
and the exponents $s$. 
The parameter $\beta$ is set to 2.0 in Fig.~\ref{figs:ratio:SMA1}(a) 
and 1.5 in Fig.~\ref{figs:ratio:SMA1}(b).
Cross represents data points of $s_{\mathrm{min}}/s$ and 
broken line indicates ${\cal G}_{\mathrm{min}}/s$.
Asterisk represents data points of $s$.
As seen in Figs.~\ref{figs:exponent_for_various_alpha} and \ref{figs::resonance:exponents:}, 
noise influences the values. 
Thus it is likely that the ratio $s_{\mathrm{min}}/s$ fluctuates and 
that the values $s_{\mathrm{min}}/s$ around the maximum of $s$ 
are below the value ${\cal G}_{\mathrm{min}}/s$.
Thus I calculate also the simple moving average of the exponents, 
and attempt to find the minimum value of them.
I take the  average of $n$ adjoining exponents and 
denote this average as $s_{\mathrm{min}}^{\mathrm{SMA}n}$.
For example, the minimum of the averages of three adjoining exponents is represented as
$s_{\mathrm{min}}^{\mathrm{SMA3}}$.
Figure~\ref{figs:ratio:SMA3} displays the $s_{\mathrm{min}}^{\mathrm{SMA3}}/s$ for $s \ge 0.3$
and the exponents $s$.
The parameter  $\beta$ is set to 2.0 in Fig.~\ref{figs:ratio:SMA3}(a) 
and 1.5 in Fig.~\ref{figs:ratio:SMA3}(b).
The symbols in Figs.~\ref{figs:ratio:SMA3}(a) and \ref{figs:ratio:SMA3}(b) 
are the same as in Figs.~\ref{figs:ratio:SMA1}(a) and \ref{figs:ratio:SMA1}(b).
It is found from 
Figs.~\ref{figs:ratio:SMA1} and \ref{figs:ratio:SMA3} that 
${\cal G}_{\mathrm{min}}/s$ is close to the values estimated by numerical calculations
around the peaks of $s$ in the resonance regions.

\section{Discussion and Conclusion}
\label{sec:conclusions}
I studied the growth of the amplitude in the Mathieu-like equation 
with multiplicative white noise.
The approximate value of the exponent at the extremum was obtained 
by introducing the width of time interval
on parametric resonance regions where parametric resonance occurs when no noise exists.
The exponents were calculated by solving the stochastic differential equations numerically 
by the symplectic numerical method.
The intensity of noise and the strength of the coupling between the noise and the variable
are reflected to the value of the parameter $\alpha$.
The value of $\alpha$ was restricted not to be negative in the present equation,
without loss of generality.
The behavior of the exponents as a function of $\alpha$ was shown roughly.

With regard to the effects of multiplicative white noise on the growth,
the band structure of the Mathieu equation is destroyed when noise exists.
The resonance structure survives for small values of $\alpha$,
and this structure is lost for large values of $\alpha$.
In the previous paper \cite{Ishihara8}, 
I investigated 
the growth in a stochastic differential equation without a periodic coefficient, 
and found that the exponent is a monotone increasing function of $\alpha$.
In contrast, 
the exponent as a function of $\alpha$ has one minimum 
on the parametric resonance region of $\alpha=0$.
This indicates the suppression of the growth by multiplicative white noise,
and this suppression occurs when the value of $\alpha$ is appropriate.
Equation~\eqref{eqn:res:exponent} can roughly explain 
the behavior of the exponent as a function of $\alpha$:
The exponent decreases with $\alpha$, reaches the minimum and increases after that.

One expects that the exponent as a function of the intensity of noise
has one minimum intuitively.
However the exponent may have some minima caused by noise. 
Theoretical expression given by Eq.~\eqref{eqn:condition_of_minimum} 
indicates that only one minimum exists. 
This fact is numerically supported too. 
It is shown theoretically and numerically in the previous sections that 
the exponent as a function of $\alpha$ has one minimum.

The minimum value of the exponent as a function of $\alpha$ was estimated 
from the numerical calculations.
I calculated the ratio $s_{\mathrm{min}}/s$: the minimum value 
divided by the exponent at $\alpha=0$, $s$.
This ratio obtained numerically is in rough agreement with that obtained theoretically
around the peaks of $s$ on the resonance regions.
The minimum value of the exponent is approximately proportional to 
the exponent $s$. 
The relative variation is of the order of 90\%,
as shown in the figures and Eq.~\eqref{eqn:min_val_of_G}.
It seems that the variation is small. 
Nevertheless, the amplitude is affected, because this is the variation of the exponent.

The decrease of Lyapunov exponent by noise was found 
in the system of an inverted Duffing oscillator with noise \cite{Mallick2004}.
The mechanism of the growth in the present case 
is different from that in the case of the inverted Duffing oscillator, 
when noise is absent.
However, the mechanism of the suppression is surely the same.
In both cases, the growth is suppressed by noise 
when the intensity of noise is appropriate.
The exponent decreases with the intensity, and reaches the minimum.
After that, the exponent increases with the intensity. 
The decrease of the exponent by white noise implies the possibility of the large 
decrease by colored noise. 
The system in which parametric resonance occurs may be stabilized by colored noise, 
as found in the system of the inverted Duffing oscillator. 

The expression of the exponent obtained theoretically includes 
the artificial parameter $\Delta z$.
Then the value of $\alpha$ at the minimum of ${\cal G}$ depends on $\Delta z$, 
while the minimum value of ${\cal G}$ is independent of $\Delta z$.
I would like to solve this problem in the future study.

\bibliographystyle{unsrt}
\bibliography{myrefs}

\end{document}